\documentclass[11pt,a4paper]{article}

\usepackage{graphicx,subfigure}  
\usepackage{color}
\usepackage{colortbl}
\graphicspath{{./figs/}} 

\usepackage{amsmath} 
\usepackage{amssymb}
\usepackage{amsfonts}
\usepackage{xspace}
\usepackage{ifthen} 

\newboolean{articletitles}
\setboolean{articletitles}{true} 

\newboolean{uprightparticles}
\setboolean{uprightparticles}{false} 

\newboolean{inbibliography}
\setboolean{inbibliography}{false} 

\usepackage{mciteplus}
\usepackage{lineno}

\usepackage{xspace} 
\usepackage{upgreek}

\def\lhcb {\mbox{LHCb}\xspace}

\def\MagUp {\mbox{\em Mag\kern -0.05em Up}\xspace}

\ifthenelse{\boolean{uprightparticles}}%
{

 \def\Pmu         {\ensuremath{\upmu}\xspace}

 \def\Ppi         {\ensuremath{\uppi}\xspace}

 \def\Ppsi        {\ensuremath{\uppsi}\xspace}

 \def\PDelta      {\ensuremath{\Delta}\xspace}                 
 \def\PXi      {\ensuremath{\Xi}\xspace}                 
 \def\PLambda      {\ensuremath{\Lambda}\xspace}                 
 \def\PSigma      {\ensuremath{\Sigma}\xspace}                 
 \def\POmega      {\ensuremath{\Omega}\xspace}                 
 \def\PUpsilon      {\ensuremath{\Upsilon}\xspace}                 
 
 \def\PB      {\ensuremath{\mathrm{B}}\xspace}                 
                  
 \def\PD      {\ensuremath{\mathrm{D}}\xspace}

 \def\PJ      {\ensuremath{\mathrm{J}}\xspace}                 
 \def\PK      {\ensuremath{\mathrm{K}}\xspace}

 \def\Pb      {\ensuremath{\mathrm{b}}\xspace}

 \def\Pi      {\ensuremath{\mathrm{i}}\xspace}

 \def\Ps      {\ensuremath{\mathrm{s}}\xspace}

}
{

 \def\Pmu         {\ensuremath{\mu}\xspace}

 \def\Ppi         {\ensuremath{\pi}\xspace}

 \def\Ppsi        {\ensuremath{\psi}\xspace}                 
                  
 \mathchardef\PDelta="7101
 \mathchardef\PXi="7104
 \mathchardef\PLambda="7103
 \mathchardef\PSigma="7106
 \mathchardef\POmega="710A
 \mathchardef\PUpsilon="7107
                  
 \def\PB      {\ensuremath{B}\xspace}                 
                  
 \def\PD      {\ensuremath{D}\xspace}

 \def\PJ      {\ensuremath{J}\xspace}                 
 \def\PK      {\ensuremath{K}\xspace}

 \def\Pb      {\ensuremath{b}\xspace}

 \def\Pi      {\ensuremath{i}\xspace}

 \def\Ps      {\ensuremath{s}\xspace}

}

\makeatletter
\ifcase \@ptsize \relax
  \newcommand{\miniscule}{\@setfontsize\miniscule{4}{5}}
\or
  \newcommand{\miniscule}{\@setfontsize\miniscule{5}{6}}
\or
  \newcommand{\miniscule}{\@setfontsize\miniscule{5}{6}}
\fi
\makeatother

\DeclareRobustCommand{\optbar}[1]{\shortstack{{\miniscule (\rule[.5ex]{1.25em}{.18mm})}
  \\ [-.7ex] $#1$}}

\def\mup        {{\ensuremath{\Pmu^+}}\xspace}
\def\mun        {{\ensuremath{\Pmu^-}}\xspace} 

\def\squark    {{\ensuremath{\Ps}}\xspace}

\def\bquark    {{\ensuremath{\Pb}}\xspace}

\def\pion   {{\ensuremath{\Ppi}}\xspace}

\def\pip    {{\ensuremath{\pion^+}}\xspace}
\def\pim    {{\ensuremath{\pion^-}}\xspace}

\def\kaon    {{\ensuremath{\PK}}\xspace}
  \def\Kbar    {{\kern 0.2em\overline{\kern -0.2em \PK}{}}\xspace}

\def\KorKbar    {\kern 0.18em\optbar{\kern -0.18em K}{}\xspace}

\def\Kp      {{\ensuremath{\kaon^+}}\xspace}
\def\Km      {{\ensuremath{\kaon^-}}\xspace}
\def\Kpm     {{\ensuremath{\kaon^\pm}}\xspace}

\def\KS      {{\ensuremath{\kaon^0_{\mathrm{ \scriptscriptstyle S}}}}\xspace}

  \def\Dbar    {{\kern 0.2em\overline{\kern -0.2em \PD}{}}\xspace}

\def\DorDbar    {\kern 0.18em\optbar{\kern -0.18em D}{}\xspace}

\def\B       {{\ensuremath{\PB}}\xspace}
\def\Bbar    {{\ensuremath{\kern 0.18em\overline{\kern -0.18em \PB}{}}}\xspace}

\def\BorBbar    {\kern 0.18em\optbar{\kern -0.18em B}{}\xspace}
\def\Bz      {{\ensuremath{\B^0}}\xspace}
\def\Bzb     {{\ensuremath{\Bbar{}^0}}\xspace}

\def\Bpm     {{\ensuremath{\B^\pm}}\xspace}

\def\Bd      {{\ensuremath{\B^0}}\xspace}
\def\Bs      {{\ensuremath{\B^0_\squark}}\xspace}
\def\Bsb     {{\ensuremath{\Bbar{}^0_\squark}}\xspace}

\def\jpsi     {{\ensuremath{{\PJ\mskip -3mu/\mskip -2mu\Ppsi\mskip 2mu}}}\xspace}
\def\psitwos  {{\ensuremath{\Ppsi{(2S)}}}\xspace}

  \def\Y#1S{\ensuremath{\PUpsilon{(#1S)}}\xspace}

\def\Lz          {{\ensuremath{\PLambda}}\xspace}
\def\Lbar        {{\ensuremath{\kern 0.1em\overline{\kern -0.1em\PLambda}}}\xspace}
\def\LorLbar    {\kern 0.18em\optbar{\kern -0.18em \PLambda}{}\xspace}

\def\Lb      {{\ensuremath{\Lz^0_\bquark}}\xspace}

\def\to                 {\ensuremath{\rightarrow}\xspace}

\def\CP                {{\ensuremath{C\!P}}\xspace}

\def\AT#1     {\ensuremath{A_{\mathrm{T}}^{#1}}\xspace}           

\def\C#1      {\ensuremath{\mathcal{C}_{#1}}\xspace}                       
\def\Cp#1     {\ensuremath{\mathcal{C}_{#1}^{'}}\xspace}                    
\def\Ceff#1   {\ensuremath{\mathcal{C}_{#1}^{\mathrm{(eff)}}}\xspace}        
\def\Cpeff#1  {\ensuremath{\mathcal{C}_{#1}^{'\mathrm{(eff)}}}\xspace}       
\def\Ope#1    {\ensuremath{\mathcal{O}_{#1}}\xspace}                       
\def\Opep#1   {\ensuremath{\mathcal{O}_{#1}^{'}}\xspace}                    

\newcommand{\tev}{\ifthenelse{\boolean{inbibliography}}{\ensuremath{~T\kern -0.05em eV}}{\ensuremath{\mathrm{\,Te\kern -0.1em V}}}\xspace}
\newcommand{\gev}{\ensuremath{\mathrm{\,Ge\kern -0.1em V}}\xspace}
\newcommand{\mev}{\ensuremath{\mathrm{\,Me\kern -0.1em V}}\xspace}
\newcommand{\kev}{\ensuremath{\mathrm{\,ke\kern -0.1em V}}\xspace}
\newcommand{\ev}{\ensuremath{\mathrm{\,e\kern -0.1em V}}\xspace}
\newcommand{\gevc}{\ensuremath{{\mathrm{\,Ge\kern -0.1em V\!/}c}}\xspace}
\newcommand{\mevc}{\ensuremath{{\mathrm{\,Me\kern -0.1em V\!/}c}}\xspace}
\newcommand{\gevcc}{\ensuremath{{\mathrm{\,Ge\kern -0.1em V\!/}c^2}}\xspace}
\newcommand{\gevgevcccc}{\ensuremath{{\mathrm{\,Ge\kern -0.1em V^2\!/}c^4}}\xspace}
\newcommand{\mevcc}{\ensuremath{{\mathrm{\,Me\kern -0.1em V\!/}c^2}}\xspace}

\def\gsim{{~\raise.15em\hbox{$>$}\kern-.85em
          \lower.35em\hbox{$\sim$}~}\xspace}
\def\lsim{{~\raise.15em\hbox{$<$}\kern-.85em
          \lower.35em\hbox{$\sim$}~}\xspace}

\def\tell1  {TELL1\xspace}
\def\ukl1   {UKL1\xspace}

\begin{document}

\title{CP violation in $b$ hadrons at LHCb\\
      {\it \footnotesize Proceedings of the Particles and Nuclei International Conference 2017, Beijing, China}}
      
\author{A. Hicheur\\
  { \it On behalf of the LHCb collaboration}\\  
{ \footnotesize Instituto de F\'isica, Universidade Federal do Rio de Janeiro}\\
{ \footnotesize CEP 21945-970 Rio de Janeiro, RJ,
Brazil\footnote{also at: Department of Physics, University of Constantine 1, 325 Route de Ain El Bey, Constantine 25017, Algeria.}}\\
{ \footnotesize hicheur@if.ufrj.br}}

\maketitle


\begin{abstract}
The most recent results on \CP violation in $b$ hadrons obtained by the LHCb Collaboration with Run I and years 2015-2016 of Run II are reviewed. The different types of violation are covered by the studies presented in this paper.
\end{abstract}

\section{Introduction}
The violation of the \CP symmetry is predicted in the Standard Model (SM) weak sector and closely related to the existence of at least three flavour families. However, other sources of \CP violation are likely to intervene to explain the excess of matter over antimatter in the universe.
For what concerns the quarks, the SM-based \CP violating phase is embedded in the Cabibbo-Kobayashi-Maskawa (CKM) matrix \cite{CKM1,CKM2}: 
\begin{equation}
\label{CKM_matrix}
 V_{CKM}  =
\left[ 
       \begin{array}{c c c} 
              V_{ud} &  V_{us} & V_{ub} \\
              V_{cd} &  V_{cs} & V_{cb} \\
    	      V_{td} &  V_{ts} & V_{tb} 
       \end{array} 
\right]      
\end{equation}
involving couplings within and between quark generations, because the eigenstates of the weak interaction are different from the mass eigenstates. The magnitude of the CKM elements is not evenly distributed, though: the modules of the diagonal elements (within-generation couplings) are one while the strength decreases as one departs from the diagonal.  The weakest couplings are thus between the first and third generation, namely $V_{td}$ and $V_{ub}$.
The unitarity condition, $ V_{CKM} V_{CKM}^\dagger=\mathbf{1}$, imposes six independent relations, of which one is of particular interest, $V_{ ud } V_{ ub }^*+V_{ cd } V_{ cb }^*+V_{ td } V_{ tb }^*=0$, involving a sum of elements of similar magnitude and defining the ``\Bd triangle'' or Unitarity Triangle , and a second one, $V_{ us} V_{ ub }^*+V_{ cs } V_{ cb }^*+V_{ ts } V_{ tb }^*=0$, although being an unbalanced squeezed triangle, involves the angle intervening in the \Bs meson oscillations, $\beta_s=\mathrm{arg}\left(\frac{-V_{ts}V_{tb}^*}{V_{cs}V_{cb}^*}\right)$.

To probe \CP violation, rate asymmetries of decays $B\to f$ are measured, involving amplitudes such as $A_f = \langle f |H_{eff}|B\rangle$, $A_{\overline f} = \langle \overline f|H_{eff}|B\rangle$, $\overline A_f = \langle f |H_{eff}|\overline B\rangle$ and $\overline A_{\overline f} = \langle \overline f |H_{eff}|\overline B\rangle$. $\overline B$ and $\overline f$ denote the charge conjugates of the $b$ hadron and the final state, respectively. $H_{eff}$ is the effective Hamiltonian describing the decay. Physics beyond SM may intervene, in box or loop diagrams, as contributions to $H_{eff}$ and thus impact directly on the asymmetries.

 LHCb research program has focused on \CP violation in the decays, in the neutral B mesons mixing, and in the interference of decays and mixing. Recent results spanning the three categories are reviewed in this paper.

\section{\CP violation in mixing}
\label{sec:mixing}
Neutral $B$ mesons oscillate between $B^0_q$ and $\overline B^0_q$ states ($q=d,s$) \footnote{$B^0_d = \Bd$ throughout the paper.}. The oscillation dynamics can be written as:
\begin{equation}
i\frac{d}{dt}\begin{pmatrix}|B^0_q(t)\rangle \vspace{1mm}\\|\overline B^0_q(t)\rangle\end{pmatrix}=\Big[\begin{pmatrix}M_{11}^q & M_{12}^q \vspace{2mm}\\ M_{12}^{q*} & M_{22}^q\end{pmatrix}-\frac{i}{2}\begin{pmatrix}\Gamma_{11}^q & \Gamma_{12}^q \vspace{2mm}\\\Gamma_{12}^{q*} & \Gamma_{22}^q\end{pmatrix}\Big]\begin{pmatrix}|B^0_q(t)\rangle \vspace{1mm}\\|\overline B^0_q(t)\rangle\end{pmatrix},
\end{equation}
where $M_{ij}^q$ and $\Gamma_{ij}^q$ are elements of the mass and decay matrices. We note $\phi^q_{12}=\mathrm{arg}(-M_{12}^q/\Gamma_{12}^q)$ and $\phi^q_{M}=\mathrm{arg}(M_{12}^q)$. The eigenstates of the Hamiltonian are noted as $|B^0_{L,H}\rangle = p|B^0\rangle \pm q|\overline B^0\rangle$, where L and H stand for light and heavy, respectively.
In the SM, \CP violation in mixing $\phi^q_{12}$ is expected to be very small (or said otherwise, $\left|\frac{q}{p}\right|\approx 1$) and thus an intervention of new heavy particles could be detected through a substantial deviation from the null prediction.\\
For both $\Bd$ and $\Bs$, semileptonic (and thus flavour-specific) decays are used to obtain
\begin{equation}
A_{sl}^q =\frac{\Gamma(\overline B^0_q \to B^0_q \to f)-\Gamma(B^0_q \to \overline B^0_q \to \overline f)}{\Gamma(\overline B^0_q \to B^0_q \to f)+\Gamma(B^0_q \to \overline B^0_q \to \overline f)} = -2 \left(\left|\frac{q}{p}\right|-1\right)
\end{equation}

Experimentally, \lhcb analyses measure the time-dependent yield asymmetry of the decay states \cite{LHCb-PAPER-2014-053,LHCb-PAPER-2016-013}:
\begin{equation}
\frac{N(B^0_q \to f,t)-N(\overline B^0_q \to \overline f,t)}{N(B^0_q \to f,t)+N(\overline B^0_q \to \overline f,t)}=A_D + \frac{A_{sl}^q}{2}-\left(A_P + \frac{A_{sl}^q}{2}\right)\frac{\cos(\Delta M_q t)}{\cosh(\Delta \Gamma_q t/2)}
\end{equation}
where $\Delta M_q$ and $\Delta\Gamma_q$ are the mass and width differences, $A_P$ and $A_D$ are the $B$-meson production asymmetry and the combined decay products detection asymmetry. For \Bs, the time-dependent term cancels due to the fast oscillations. In the \Bd case, a time-depend fit is necessary.

Using $\Bd\to D^-(\to K\pi\pi) \mu^+ \nu X$, $\Bd\to D^{*-}(\to D^0(K\pi)\pi) \mu^+ \nu X$, and $\Bs\to D_s^-(\to KK\pi) \mu^+ \nu X$ decay chains, the \lhcb measurements are \cite{LHCb-PAPER-2014-053,LHCb-PAPER-2016-013}:
\begin{center}
$A_{sl}^d =(-0.02\pm0.19(\mathrm{stat})\pm0.30(\mathrm{syst}))\%$, $A_{sl}^s =(0.39\pm0.26(\mathrm{stat})\pm0.20(\mathrm{syst}))\%$,
\end{center}
where the systematic (second) uncertainty is dominated by the size of the control samples used in the determination of the detection asymmetry. These numbers are compatible with the SM-based expectations of orders $10^{-4}$ and $10^{-5}$, respectively~\cite{artusolenz_asl}.

\section{\CP violation in decay}
\CP violation in decay occurs whenever $|A_f|^2 \neq |\overline A_{\overline f}|^2$. This requires flavor-specific $B$ decays and could happen only if $A_f$ contains at least two amplitudes with different strong $\delta_i$ and weak $\varphi_i$ phases. In that case, $|\overline A_{\overline f}|^2 - |A_f|^2 =-4|A_1||A_2|\sin(\delta_1-\delta_2)\sin(\varphi_1-\varphi_2)$. The extraction of the parameter of interest, the weak phase difference $\varphi_1-\varphi_2$, is thus limited by the knowledge of the strong phase difference. However, various fitting techniques involving ratios of branching fractions and asymmetries are used to circumvent this limitation.

\subsection{CKM $\gamma$ angle}
The angle $\gamma=\mathrm{arg}\left(\frac{-V_{ud}V_{ub}^*}{V_{cd}V_{cb}^*}\right)$ is the least constrained angle of the Unitarity Triangle and is the subject of extensive \lhcb studies involving $B\to DK$ decays \cite{LHCb-CONF-2017-004}.

The principle of the measurement relies on the use of the interference of tree favored $b\to c (\overline u s)$ ($A_B$) and suppressed $b\to u (\overline c s)$ ($A_Br_B e^{i(\delta_B-\gamma)}$, $r_B$ is the suppression factor and $\delta_B$ is the relative strong phase) decays leading to final states of the type $B\to D X_s$ where $D$ is a charm meson and $X_s$ a strange system ($K$, $K^*$, $K\pi$, $K\pi\pi\pi$, ...). The charm meson decays to a final state accessible by both $D$ and $\overline D$ through allowed ($A_D$) and suppressed ($A_Dr_De^{i\delta_D}$) amplitudes. Since more than two decades, several methods have initially been proposed for $B^+ \to D K^+$ \cite{GLW,ADS,GGSZ,GLS} to extract the angle for observables including charge asymmetries and ratio of branching fractions which definition may vary depending on the type of $D$ decay (\CP eigenstate, flavour eigenstate, multibody). Beside the generalization of the paradigm to final states involving $D^*$ or more complex $X_s$ system, alternative methods have been developped with other $B$ mesons, such as $B^0_s \to D^\mp_s K^\pm$ \cite{LHCb-CONF-2016-015}. The most recent \lhcb combination of measurements \cite{LHCb-CONF-2017-004} gives an average of $\gamma = (76.8^{+5.1}_{-5.7})^\circ$, which dominates the world average, $(76.2^{+4.7}_{-5.0})^\circ$ \cite{HFAG}.

\subsection{\CP violation in baryonic decays}
\begin{figure}[t]
\begin{center}
\includegraphics[width=0.45\textwidth]{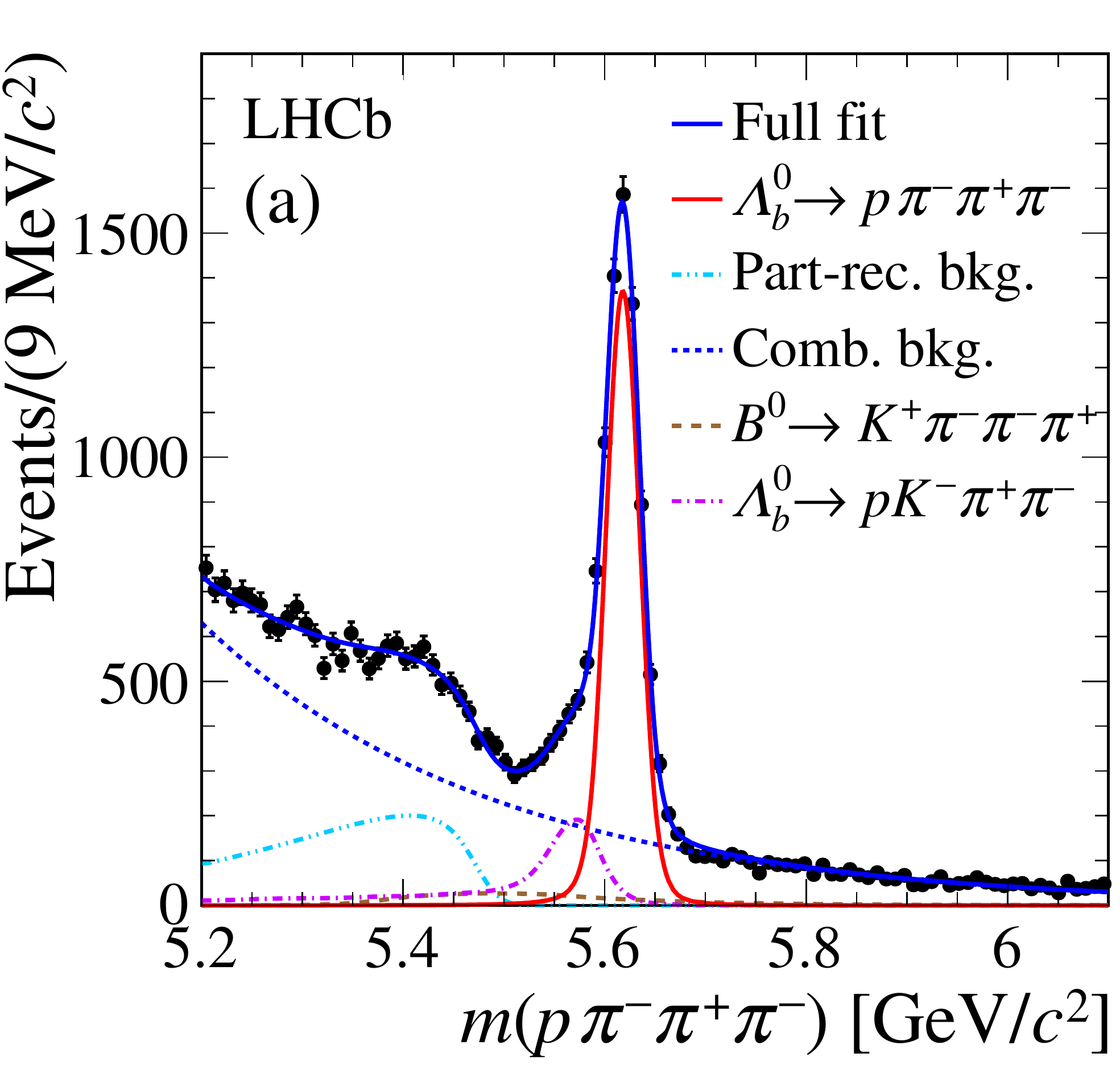}
\includegraphics[width=0.45\textwidth]{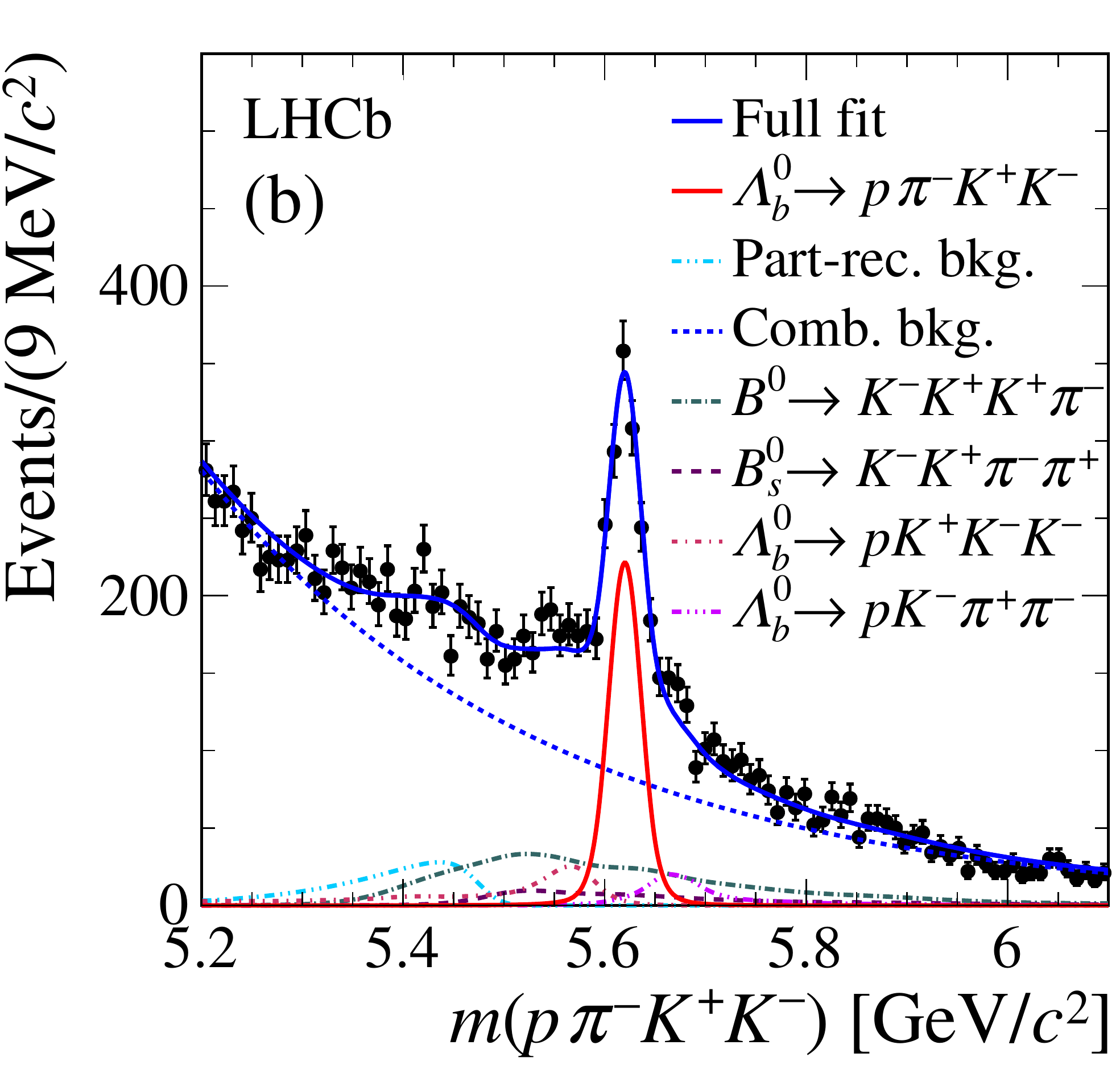}
\end{center}
\caption{Invariant mass spectrum for (left) $p\pim \pi^+\pi^-$ and (right) $p\pim K^+K^-$.}
\label{Fig:Lbtoppihh_mass}
\end{figure}

A pioneering study has been performed on four-body decays of the \Lb baryon, $\Lb\to p\pim h^+h^-$ ($h=\pi,K$) \cite{LHCb-PAPER-2016-030}. For four-body modes, a useful observable which has been chosen for \CP studies is the triple scalar product $C_{\hat{T}}=\vec{p}_p.(\vec{p}_{h_1^-}\times\vec{p}_{h_2^+})$ \cite{triplscalar1}. $h_1=\pi$ and $h_2=K$ for $\Lb\to p\pim K^+K^-$, and $h_1=h_2=\pi$ for $\Lb\to p\pim \pi^+\pi^-$. In the latter case, the ambiguity for the choice of $h_1$ is removed by selecting the fastest $\pi^-$ in the \Lb rest frame, $\pi_{fast}^-$. \CP conjugation applied to $C_{\hat{T}}$ gives $-\vec{p}_{\overline p}.(\vec{p}_{h_1^+}\times\vec{p}_{h_2^-}) = -\overline{C}_{\hat{T}}$. Therefore, for \CP symmetry to be conserved, any observable depending on $C_{\hat{T}}$ should be unchanged under the substitution $C_{\hat{T}}\to -\overline{C}_{\hat{T}}$. Experimentally, the following yield asymmetries have been chosen:  $A_{\hat{T}}(C_{\hat{T}})=\frac{N(C_{\hat{T}}>0)-N(C_{\hat{T}}<0)}{N(C_{\hat{T}}>0)+N(C_{\hat{T}}<0)}$, $\overline{A}_{\hat{T}}(\overline{C}_{\hat{T}})=\frac{N(-\overline{C}_{\hat{T}}>0)-N(-\overline{C}_{\hat{T}}<0)}{N(-\overline{C}_{\hat{T}}>0)+N(-\overline{C}_{\hat{T}}<0)}$. Defining the difference, $a_{CP}^{{\hat{T}\mbox{-}\mathrm{odd}}}=\frac{1}{2}(A_{\hat{T}}-\overline{A}_{\hat{T}})$ \cite{triplscalar2}, the test consists in probing deviations from zero of $a_{CP}^{{\hat{T}\mbox{-}\mathrm{odd}}}$.
Figure \ref{Fig:Lbtoppihh_mass} shows the invariant mass spectra of the $p\pim h^+h^-$ systems. Both $\Lb\to p\pim h^+h^-$ signals are observed for the first time, with yields of $6646 \pm 105$ ($h=\pi$) and $1030 \pm 56$ ($h=K$) events.

\begin{figure}[t]
\begin{center}
\includegraphics[width=0.45\textwidth]{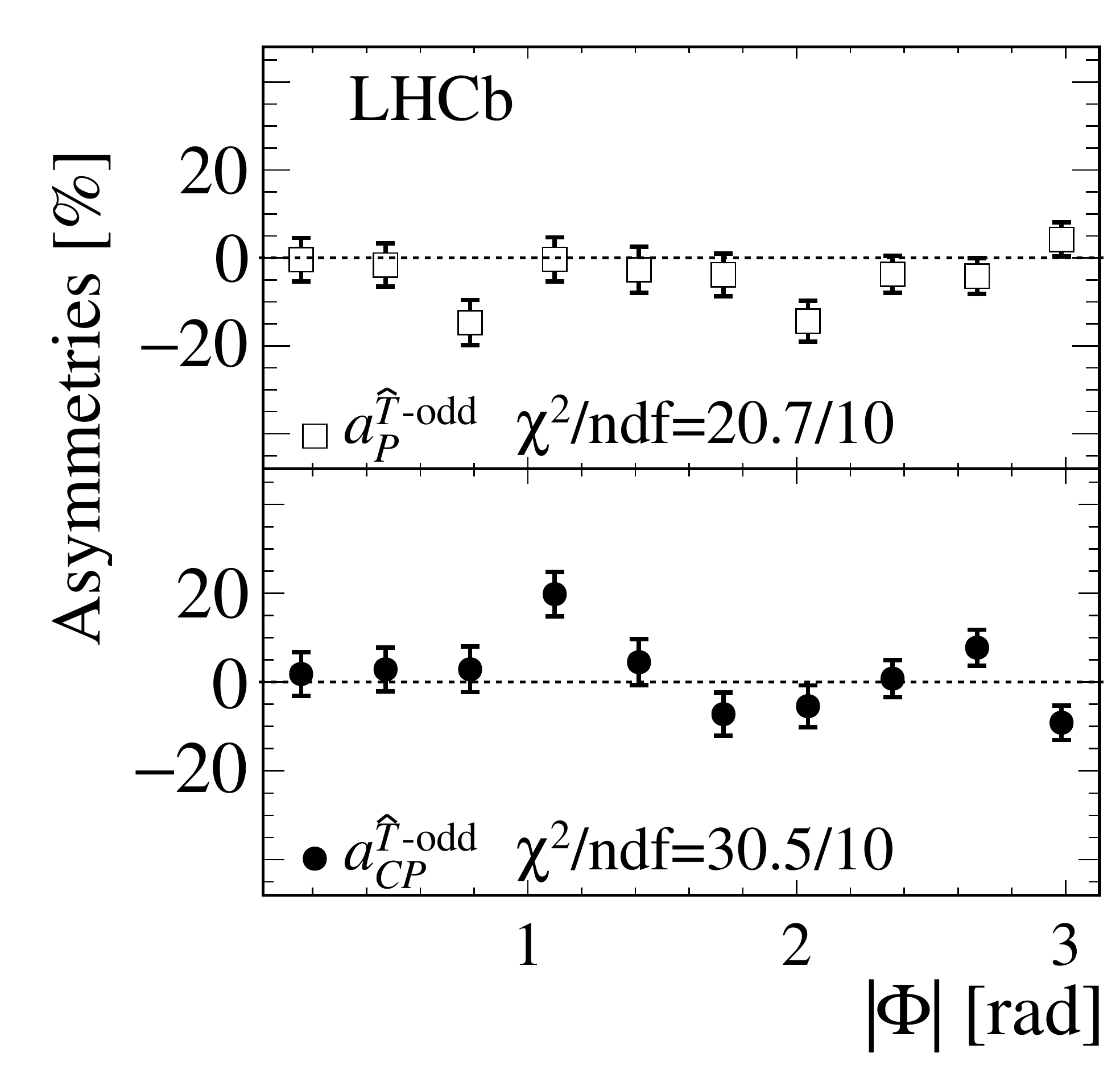}
\end{center}
\caption{Variation of (top) $a_{P}^{{\hat{T}\mbox{-}\mathrm{odd}}}=\frac{1}{2}(A_{\hat{T}}+\overline{A}_{\hat{T}})$ and (bottom) $a_{CP}^{{\hat{T}\mbox{-}\mathrm{odd}}}$ as a function of $|\Phi|$ for $\Lb\to p\pim \pi^+\pi^-$ .}
\label{Fig:acpodd_schemeB}
\end{figure}

$a_{CP}^{{\hat{T}\mbox{-}\mathrm{odd}}}$ is then studied as a function of the angle $\Phi$ between the planes formed by $(p,K^-(\pi_{fast}^-))$ and $(\pi^-_{(slow)},K^+)$, where we note that $C_{\hat{T}}\propto \sin(\Phi)$. In the case of $\Lb\to p\pim \pi^+\pi^-$, substantial deviation from zero is observed, as illustrated in Fig.\ref{Fig:acpodd_schemeB}, with a combined significance of 3.3$\sigma$. This is the first evidence of \CP violation in the decay of a baryon. With the current statistics, no significant deviation is seen for $\Lb\to p\pim K^+K^-$.

\section{\CP violation in interferences between mixing and decay}
When neutral $B$ mesons decay to a eigen \CP final state, the possibility of oscillation implies that the measured asymmetry probing the interference of mixing and decay must be studied as a function of the decay time:
\begin{equation}
A_{CP}(t) = \frac{\Gamma(B^0_q(t) \to f_{CP})-\Gamma(\overline B^0_q(t) \to f_{CP})}{\Gamma(B^0_q(t) \to f_{CP})+\Gamma(\overline B^0_q(t) \to f_{CP})}= \frac{S_f \sin(\Delta M_q t)-C_f \cos(\Delta M_q t)}{\cosh(\Delta \Gamma_q t/2)+A_f^{\Delta\Gamma}\sinh(\Delta \Gamma_q t/2)}
\end{equation}
where $C_f = \frac{1-|\lambda_f|^2}{1+|\lambda_f|^2}$, $S_f = \frac{2\Im(\lambda_f)}{1+|\lambda_f|^2}$, $A_f^{\Delta\Gamma}= - \frac{2\Re(\lambda_f)}{1+|\lambda_f|^2}$ and $\lambda_f=\frac{q}{p}\frac{\overline A_f}{A_f}$.
For the \Bd meson, the denominator is equal to 1, since $\Delta\Gamma_d/\Gamma_d << 1$. Following the discussion in section \ref{sec:mixing}, $\left|\frac{q}{p}\right|=1$. If only one amplitude contributes to the decay, $\lambda_f=e^{i\Phi_q}=e^{i(\Phi_M-2\Phi_D)}$ (and thus $C_f =0$), where $\Phi_M$ and $\Phi_D$ are the weak mixing and decay phases, respectively. For both \Bd and \Bs, one can reasonably assumes that $\Phi_M \simeq 0$. Experimentally, except for the mixing phase simplification, the quantities $S_f$, $C_f$, and $\Delta M_q$ (and $A_f^{\Delta\Gamma}$, $\Delta \Gamma_q$ for \Bs) are extracted without prior assumption.
\subsection{$\beta_s$ and $\beta$ measurements}
The $b\to c\overline c s$ tree decays are the reference modes for extracting the CKM angles $\beta=arg\left(-\frac{V_{cd}V_{cb}^*}{V_{td}V_{tb}^*}\right)$ ($\Phi_d=-2\beta$ for \Bd) and $\beta_s=arg\left(-\frac{V_{ts}V_{tb}^*}{V_{cs}V_{cb}^*}\right)$ ($\Phi_s=-2\beta_s$ for \Bs).

Table \ref{Tab:phis} shows all the modes used over the past years and the corresponding measurements for $\Phi_s$. The most recent one, relying on the decay $\Bs\to\jpsi\Kp\Km$ in the region $m(\Kp\Km)>m_\phi$, is the first with a tensor resonance ($f_2(1525)$) dominating the spectrum. With the current statistics, all the measurements agree with recent SM-based fits \cite{CKM_Fitter}, $-0.0365^{+0.0013}_{-0.0012}$ .

\begin{table}[!htb]
\caption{\lhcb $\Phi_s$ measurements, with their respective statistical (first) and systematic (second) uncertainties.}
    {\begin{tabular}{ccc}
        \hline
        Final state & $\Phi_s$ (rad) & Reference\\
        \hline
        $\jpsi\pip\pim$ (including $f_0$) & $+0.070\pm0.068\pm0.008$ & Phys.Lett.B B736 186 (2014) \\
        $D_s^+D_s^-$ & $+0.02\pm0.17\pm0.02$ & Phys.Rev.Lett.113 211801 (2014)\\
        $\jpsi\Kp\Km$ (including $\phi$) & $-0.058\pm0.049\pm0.006$ & Phys.Rev.Lett.114 041802 (2015)\\
        $\psitwos\phi$ & $+0.23^{+0.29}_{-0.28}\pm0.02$ & Phys.Lett.B B762, 252-262 (2016)\\
        $\jpsi\Kp\Km$ (above $\phi$) & $+0.119\pm0.107\pm0.034$ & JHEP08 (2017) 037\\
        \hline
      \end{tabular} \label{Tab:phis}}
\end{table}

For $\beta$, the Run 1 data \lhcb measurement with the golden mode $\Bd\to\jpsi(\to \mup\mun)\KS$, $\sin(2\beta)=0.731\pm0.035(\mathrm{stat})\pm0.020(\mathrm{syst})$ \cite{LHCb-PAPER-2015-004} has been recently completed by measurements based on the decays $\Bd\to\jpsi(\to e^+e^-)\KS$ and $\Bd\to\psitwos(\to \mup\mun)\KS$ \cite{LHCb-PAPER-2017-029}, resulting in a combined average of $\sin(2\beta)=0.760\pm0.034$, still in agreement with predictions\cite{CKM_Fitter}, $0.771^{+0.017}_{-0.041}$ .

\subsection{Two body $B^0_{(s)}\to hh$}
\begin{figure}[t]
\begin{center}
\includegraphics[width=0.49\textwidth]{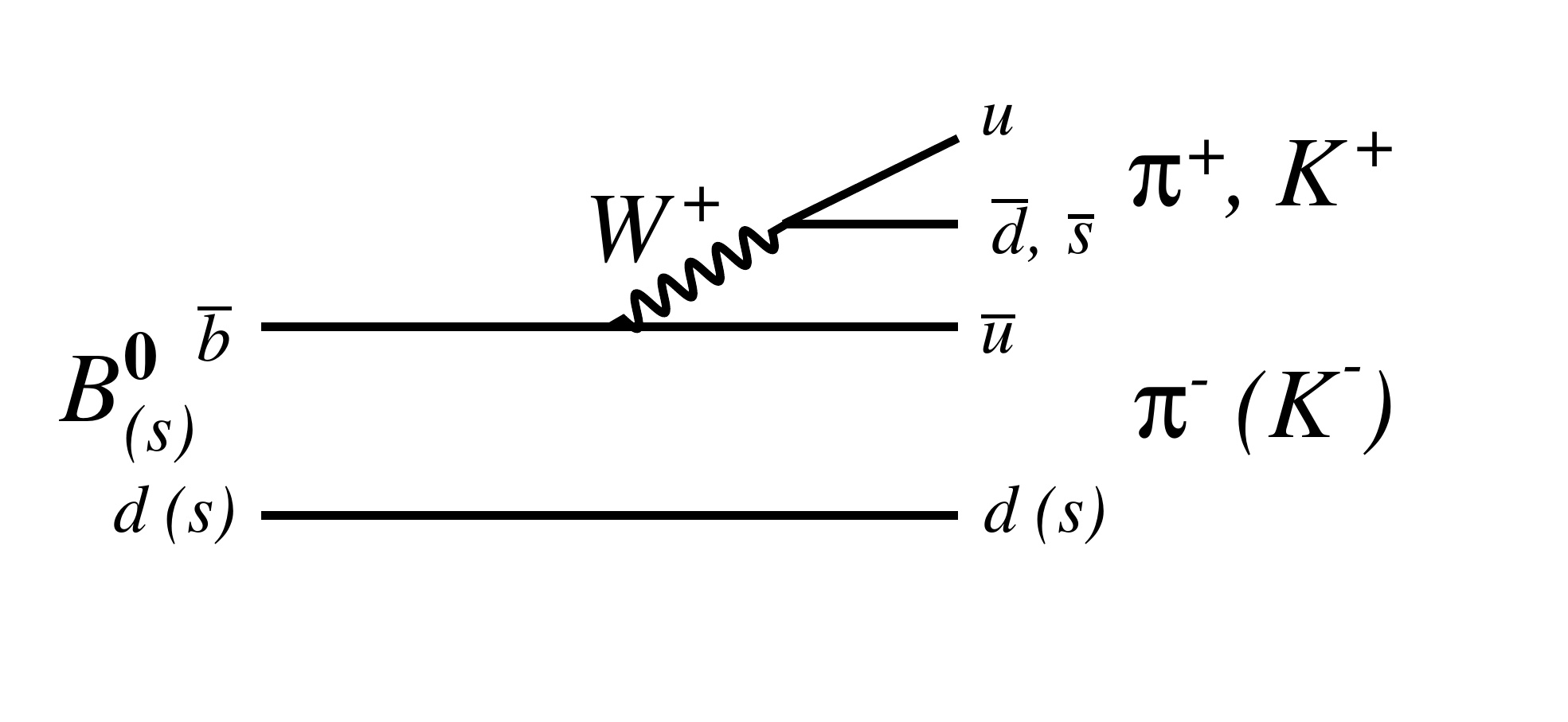}
\includegraphics[width=0.49\textwidth]{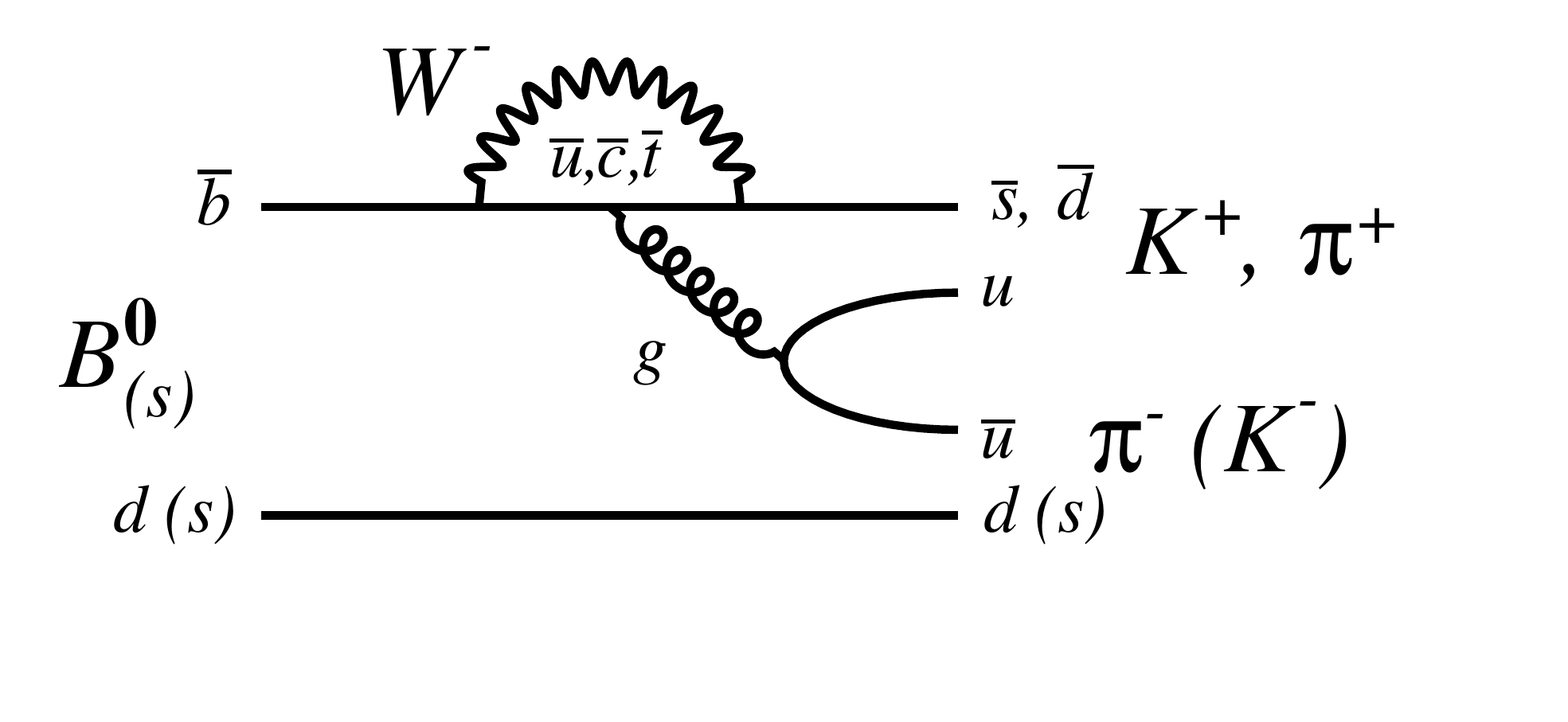}
\end{center}
\caption{Quark tree and loop (``penguin'') diagrams of the $B^0_{(s)}\to hh$ decays.}
\label{Fig:b0tohh_diag}
\end{figure}

The two body $B^0_{(s)}$ decays involve tree and loop diagram contributions, Fig.~\ref{Fig:b0tohh_diag}.
The extraction of any weak phase, here $2\beta_s$ or $\gamma$, requires some assumptions, such as U-spin symmetry \cite{Fleischer}, which inexactness limits the accuracy of the results. A recent simultaneous fit of the four channels $\Bd\to\pip\pim$, $\Bd\to\Kp\pim$, $\Bs\to\Kp\Km$, and $\Bs\to\Kp\pim$ \cite{LHCb-CONF-2016-018} led to the measurements: $C_{\pi\pi}=-0.243\pm 0.069$, $S_{\pi\pi}=-0.681\pm 0.060$, $C_{KK}=0.236\pm 0.062$, $S_{KK}=0.216\pm 0.062$ and $A_{KK}^{\Delta\Gamma}=-0.751\pm0.075$. The notable large values of the $C_f$ coefficients are due to a sizeable \CP violation in the decay produced by the interference of tree and loop diagrams. Since the significance for $(C_{KK},S_{KK},A_{KK}^{\Delta\Gamma})$ to differ from $(0,0,1)$ is 4.7$\sigma$, \lhcb establishes strong evidence for \CP violation in $\Bs\to\Kp\Km$.
\section{Summary}
Important progress has been done in the measurement of \CP violation at \lhcb with Run I and beginning of Run II data, leading to a better constraint of the Unitarity Triangle and opening new routes such as \CP violation in Baryon decays. But the needed accuracies, for a sensible comparison between indirect and direct determinations of the angles, will only be reached following the \lhcb Phase I upgrade during the next decade. As an illustration, the expected Phase I upgrade accuracy for the angle $\gamma$ is $\sim 1^\circ$, while the indirect constraint, based on loop-level processes, gives $\gamma=(66.9^{+0.94}_{-3.44})^\circ$ \cite{CKM_Fitter}. A clear indication of New Physis may show up at this stage.
\ifx\mcitethebibliography\mciteundefinedmacro
\PackageError{LHCb.bst}{mciteplus.sty has not been loaded}
{This bibstyle requires the use of the mciteplus package.}\fi
\providecommand{\href}[2]{#2}

\end{document}